\documentclass{WileyMSP-template}

\begin{document}


\title{A mono-material Nernst thermopile with hermaphroditic legs}

\maketitle


\author{Xiaokang Li*}
\author{Zengwei Zhu*}
\author{Kamran Behnia}



\begin{affiliations}
Dr. X. Li, Prof. Z. Zhu\\
Wuhan National High Magnetic Field Center and School of Physics, Huazhong University of Science and Technology, Wuhan 430074, China\\
Email Address: lixiaokang@hust.edu.cn; zengwei.zhu@hust.edu.cn

Prof. K. Behnia\\
Laboratoire de Physique et d'Etude des Mat\'{e}riaux (CNRS-Sorbonne Universit\'e)\\ ESPCI, PSL Research University, 75005 Paris, France

\end{affiliations}



\begin{abstract}

A large transverse thermoelectric response, known as anomalous Nernst effect (ANE) has been recently observed in several topological magnets.  Building a thermopile employing this effect has been the subject of several recent propositions. Here, we design and build a thermopile with  an array of tilted adjacent crystals of Mn$_3$Sn. The design employs a single material and replaces pairs of P and N thermocouples of the traditional design with hermaphroditic legs. The design exploits the large lag angle between the applied field and the magnetization, which we attribute to the interruption of magnetic octupoles at the edge of $xy$-plane. Eliminating extrinsic contacts between legs will boost the efficiency, simplify the process and pave the way for a new generation of thermopiles.

\end{abstract}


\section{Introduction}
\quad Direct conversion of heat to electricity thanks to thermoelectric devices is an environmental friendly technology for recovery of ever-increasing waste heat and heat sensing. The traditional approach, based on using the longitudinal thermoelectric (Seebeck or magneto-Seebeck) effect, has generated a vast research activity \cite{Liu2015, He2017, He2018, Niemann2016}.  Much less attention has been paid to the transverse (Nernst) thermoelectric response emerging in presence of a magnetic field~\cite{Behnia2016}. Remarkably, the record of the lowest temperature attained by thermoelectric cooling is still held by device using the Nernst effect of a bismuth crystal in a magnetic field of 11 T~\cite{Harman1964}.

\begin{figure}[h]
  \includegraphics[width=\linewidth]{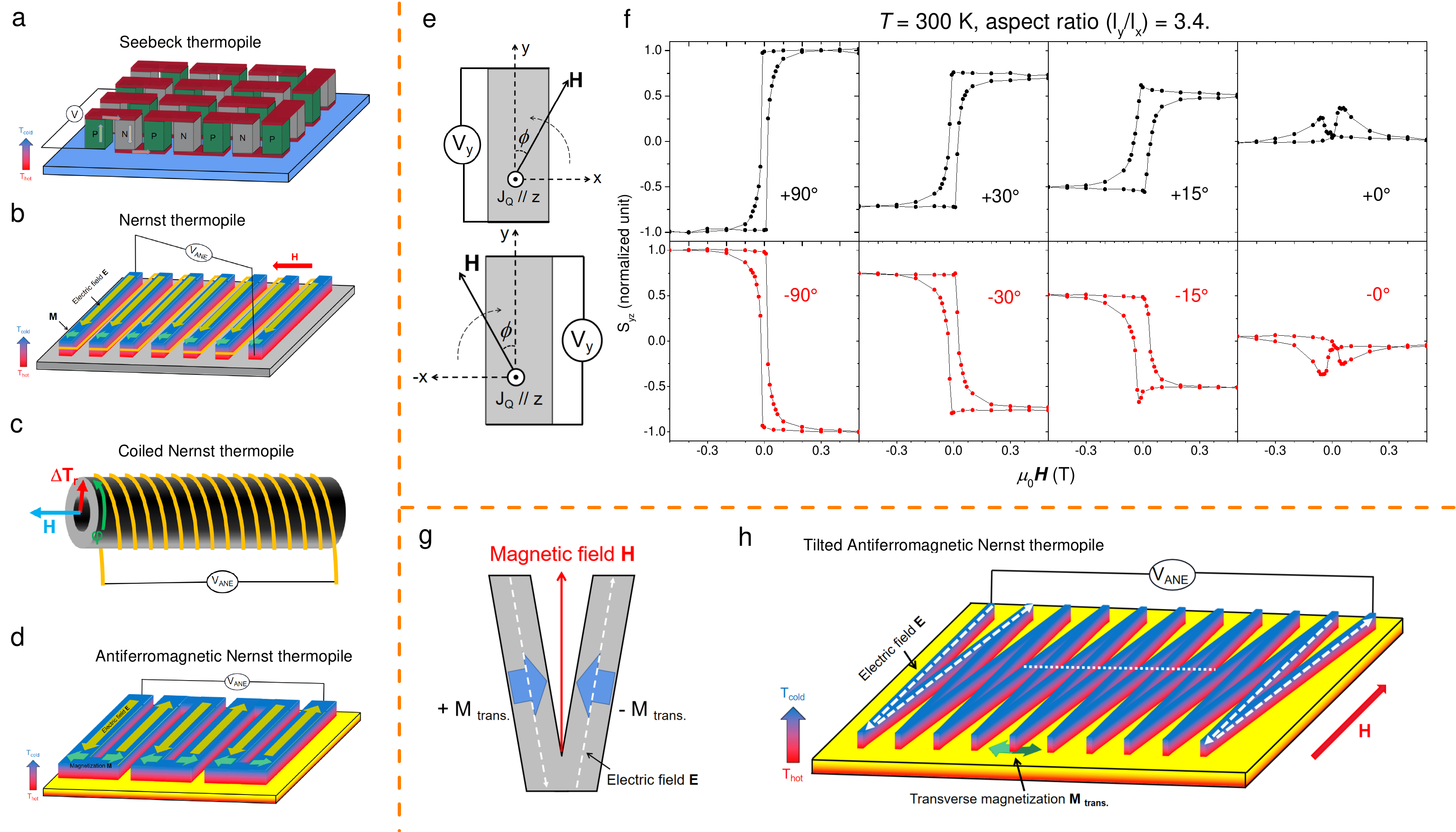}
  \caption{Previously proposed types of thermopiles and the present design: (a-d) Suggested designs for thermoelectric modules. (a) A Seebeck thermopile \cite{Liu2015, He2018} has a vertical structure and is made of two different types of materials labelled  N and P. The voltage drop is along the direction of temperature difference. (b) A Nernst thermopile \cite{Sakai2020, Sakuraba2016} exploiting the ANE has a planar structure and is suitable for thin film applications. The voltage drop is perpendicular to the direction of thermal gradient, and can be switched by the direction of magnetization. (c) A coiled Nernst thermopile \cite{ Sakuraba2016, Yang2017a, Mizuguchi2019} has a cylindrical structure with a radial temperature gradient, generating an azimuthal voltage drops along a coiled wire. (d) An antiferromagnetic Nernst thermopile \cite{Ikhlas2017} has a planar structure and is made of only one thermoelectric material, and can maintain different directions of magnetization among elements due to its tiny stray field. (e-h) Our proposed tilted Nernst thermopile (TNT) and its origin. (e) Experimental configuration: the heat current $J_{Q}$ is applied along the z-axis, the voltage $V$ is detected along y-axis and the magnetic field \textbf{$H$} rotates from $\pm$ x-axis to y-axis, in a rectangular cross-section sample with the aspect ratio $l_{y} / l_{x}$ = 3.4. (f) The Nernst responses for four different angles. The curves at $\pm 90^{\circ}$ and $\pm 0^{\circ}$ are ANE and planar Nernst effect respectively \cite{Li2019}. The data  has been normalized: $S_{ij}(\phi, H)_{normalized} = S_{ij}(\phi, H) / S_{ij}(\phi = 90^{\circ}, H = 0.5{\rm T})$. (g) A  tilted bicrystal as the elemental brick.  (h) Our proposed thermoelectric module in which we expect to automatically alternate the direction of magnetization between adjacent elements by using the transverse magnetization instead of the longitudinal magnetization.}
  \label{fig: Tilted-Nernst-thermopiles}
\end{figure}

\quad A thermopile is one component of thermoelectric technology, which converts a temperature difference to a voltage. A typical Seebeck thermopile consists of an array of linked thermocouples, each consisting of two different types of thermoelectric materials, labelled as N and P legs, and has a vertical pillar structure ( $V \, || \,\triangle T$) (See Fig.~\ref{fig: Tilted-Nernst-thermopiles}a). It requires numerous contacts between N and P legs, which dissipates the energy. Moreover, the complicated pillar structure is costly and could hamper the flexibility and endurance of the device \cite{Zhou2021}. The contact issue together with the cost and complexity of assembly process have limited the wide spread use of Seebeck thermopiles~\cite{Liu2015, He2018}.

\quad This led several authors to consider Nernst thermopiles using the anomalous Nernst effect (ANE), a transverse voltage drop perpendicular to both the magnetization and heat current, which has the potential to compensate for the shortcomings in Seebeck case \cite{Sakuraba2013, Sakai2020, Sakuraba2016, Yang2017a, Mizuguchi2019, Ikhlas2017, Zhou2020a, Fu2020}. As the thermoelectric counterpart of anomalous Hall effect (AHE), ANE has proved to be a sensitive probe of the Berry curvature near the Fermi level~\cite{Ikhlas2017, Xiao2006, Li2017, Akito2018, Guin2019, Ding2019, Xu2019, Xu2020}. In contrast to the Seebeck thermopile, a Nernst thermopile \cite{Sakai2020, Sakuraba2016} has a planar structure with $V \, \bot \,\nabla T$ (See Fig.~\ref{fig: Tilted-Nernst-thermopiles}b). The ANE signal is set by the direction of magnetization.  Fig.~\ref{fig: Tilted-Nernst-thermopiles}c shows a coil-based Nernst thermopile \cite{Sakuraba2016, Yang2017a, Mizuguchi2019}. In this case, a radial temperature gradient generates an azimuthal voltage along a coiled wire. This design has the advantage of being made of a single thermoelectric material with no contacts. However, it requires a complex axial heat source and a large permanent magnet. 

\quad Following the observation of a sizeable ANE in the noncollinear antiferromagnet Mn$_3$Sn \cite{Ikhlas2017, Li2017, Narita2017}, Ikhlas and co-workers~\cite{Ikhlas2017} proposed a new design for an antiferromagnetic Nernst thermopile with a single thermoelectric material and a simple planar heat source (Fig.~\ref{fig: Tilted-Nernst-thermopiles}d). However, this design requires alternating orientation for magnetization of adjacent elements, an outstanding problem, hitherto unresolved. 

\section{Tilted Nernst thermopile}
\quad Here, we show that the specific magnetic texture of Mn$_3$Sn allows us to circumvent this challenge and build a new type of Nernst thermopile. Mn$_3$Sn \cite{Zimmer1972, Tomiyoshi1982, Tomiyoshi1982b, Yang2017b},  a Weyl noncollinear antiferromagnet displaying a sizeable anomalous Hall effect at room temperature~\cite{Nakatsuji2015, Higo2018a}, accompanied by thermoelectric~\cite{Ikhlas2017, Li2017, Narita2017}, thermal ~\cite{Li2017} and magneto-optic ~\cite{Higo2018b} counterparts. 

\quad The property exploited in our design is illustrated in Fig.~\ref{fig: Tilted-Nernst-thermopiles}e,f, which shows the evolution of the Nernst responses in a Mn$_3$Sn sample with a rotating magnetic field. The heat current $J_{Q}$ is applied along z-axis, the voltage $V$ is measured along y-axis, and the magnetic field \textbf{$H$} rotates from along $\pm$ x-axis to along y-axis (See Fig.~\ref{fig: Tilted-Nernst-thermopiles}e).  Fig.~\ref{fig: Tilted-Nernst-thermopiles}f shows the results for different angles. The signals obtained at $\pm 90^{\circ}$ ($ 0^{\circ}$) are the anomalous (planar) Nernst effects, as reported previously~\cite{Li2019}. Interestingly, even when the magnetic field is close to the long side of the sample, with a $\pm 15^{\circ}$ misalignment, the Nernst signal is large thanks to transverse magnetization (See below for a detailed discussion). Since the sign of this signal depends on the sign of the tilted angle, the Nernst signals of two tilted crystals  (Fig.~\ref{fig: Tilted-Nernst-thermopiles}g) subject to a magnetic field aligned along their bisectrix, would add up.

\begin{figure}
  \includegraphics[width=\linewidth]{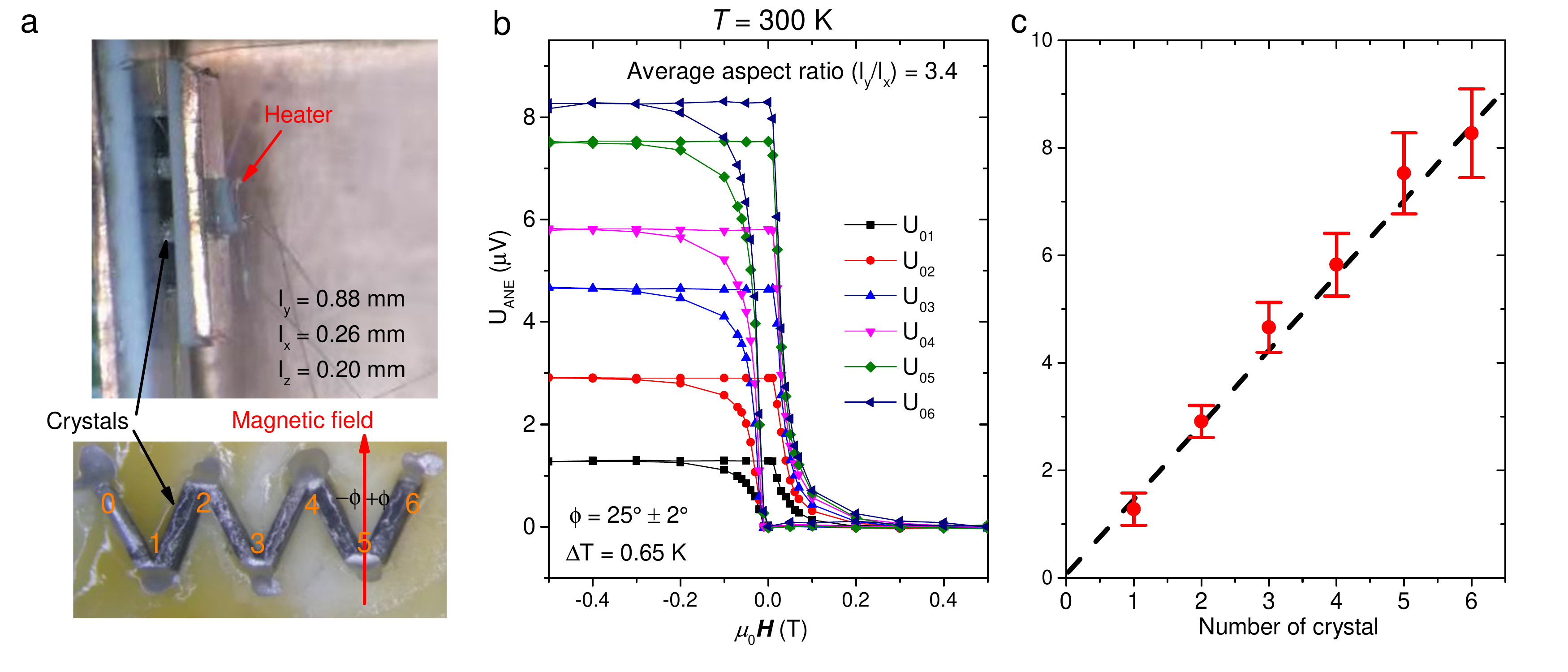}
  \caption{Proof of concept: (a) Top: Photograph of the experimental setup for the tilted Nernst thermopile. Bottom: Photograph of an array of tilted Mn$_3$Sn crystals. (b) Six different Nernst hysteresis curves, which were measured from different electrodes. (c) The extracted Nernst signal is proportional to the number of crystal. The average aspect ratio of crystals in this experiments is 3.4.}
  \label{fig:Amplify-signal}
\end{figure}

\quad This tilted bicrystal is the elemental brick of our new design for a Tilted Nernst thermopile (TNT), shown in Fig.~\ref{fig: Tilted-Nernst-thermopiles}h. We expect the voltage signals to add up linearly and boundlessly with the addition of a new elemental brick. We have realized this, as shown in Fig.~\ref{fig:Amplify-signal}.

\quad The experimental setup consists of a heater, an array of tilted crystals and a heat-sink (Fig.~\ref{fig:Amplify-signal}a). The field dependence of the Nernst voltages, measured with different electrodes from $U_{01}$ to $U_{06}$, are shown in Fig.~\ref{fig:Amplify-signal}b. One can see that the measured signal increases with increasing number of tilted crystals sandwiched between the two measuring electrodes. Extracting the signal from each curve using the formula $S_{yz}(i) = S_{yz}(i, \mu_{0}H = -0.5 {\rm T}) - S_{yz}(i, \mu_{0}H = 0.5 {\rm T})$ (Fig.~\ref{fig:Amplify-signal}c) demonstrates that, as expected, the Nernst signal increases linearly with the number of crystals.

\quad This V-shape tilted bicrystal is a hermaphroditic behaving simultaneously as N and P legs of the traditional thermopiles. Its presence eliminates the inevitable contacts requiring an extrinsic material.

\section{ The dependence of the planar Hall effect and the lag angle on the aspect ratio}
\begin{figure}
  \includegraphics[width=\linewidth]{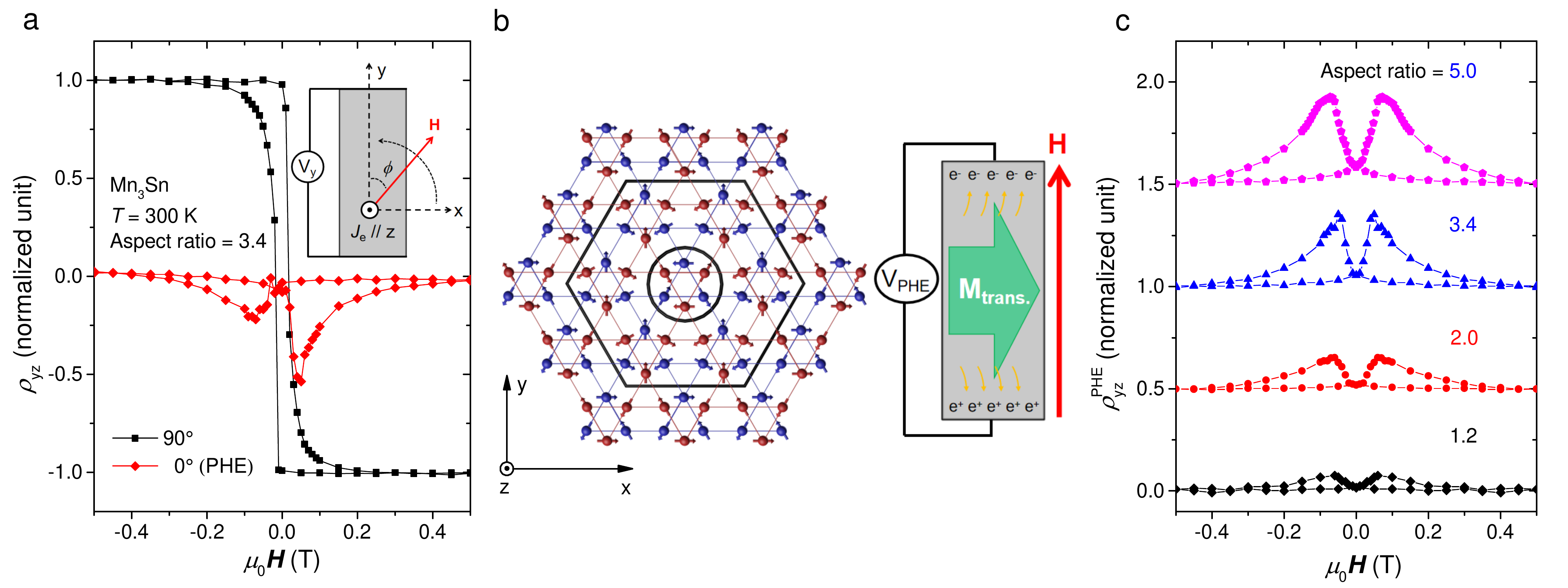}
  \caption{Planar Hall effect and its dependence on aspect ratio: (a) Anomalous Hall effect ($90^{\circ}$) and planar Hall effect ($0^{\circ}$) in a sample with the aspect ratio equals to 3.4. The inset shows the experimental configuration. The charge current $J_{e}$ is applied along the z-axis and the voltage $V$ is detected along the y-axis and the magnetic field \textbf{$H$} is in the xy-plane. The PHE refers to the signal occurring when the field is parallel to the direction of voltage. (b) A sketch of domain reversal in Mn$_3$Sn (left) and the relation between PHE and transverse magnetization (right). (c) Variation of PHE with aspect ratio. The amplitude of the signal steadily increases with the aspect ratio, implying that a large aspect ratio favors transverse magnetization. The data has been normalized: $\rho_{ij}(\phi, H)_{normalized} = \rho_{ij}(\phi, H) / \rho_{ij}(\phi = 90^{\circ}, H = 0.5{\rm T})$, and the data with the aspect ratio = 5 is from \cite{Li2019}.}
  \label{fig:PHE-TM}
\end{figure}

\quad Our TNT employs the sizeable transverse magnetization emerging during the domain reversal process in Mn$_3$Sn. A finite transverse magnetization was directly observed in a previous study employing Hall probes~\cite{Li2019} and was identified as the origin of the planar Hall effect (PHE). This is a Hall signal in which the external magnetic field is parallel but not perpendicular to the voltage direction ($H \, || \, V$). The PHE signal emerges inside the hysteresis and, as seen in Fig.~\ref{fig:PHE-TM}a, is restricted to a  narrow window where the AHE signal is changing rapidly, due to the reversal of the spins by the magnetic field. 

\quad Fig.~\ref{fig:PHE-TM}b presents a schematic sketch of spin reversal in Mn$_3$Sn. Each octupole-like spin cluster \cite{Suzuki2017}, consists of six Mn atoms and their spins. When a new domain with opposite polarity nucleates inside a previously dominant domain, spins  gradually rotate clockwise or anticlockwise from the center towards the periphery. This generates a transverse magnetization.  When electrons flow along the z-axis, they feel the Lorentz force generated by the magnetization along x-axis, will be deflected along the y-axis generating a voltage parallel to the magnetic field (Fig.~\ref{fig:PHE-TM}b). 

\quad An important feature of the PHE is that it depends on the aspect ratio  ($ar = l_{y} / l_{x}$) of the crystals. This is shown in  Fig.~\ref{fig:PHE-TM}c. As seen in the figure, the amplitude of the PHE steadily increases as $ar$ passes from 1.2 to 5.0. This implies that a large aspect ratio facilitates the formation of transverse magnetization along short side. This observation provides a crucial clue to identify the origin of the transverse magnetization. 

\begin{figure}
  \includegraphics[width=\linewidth]{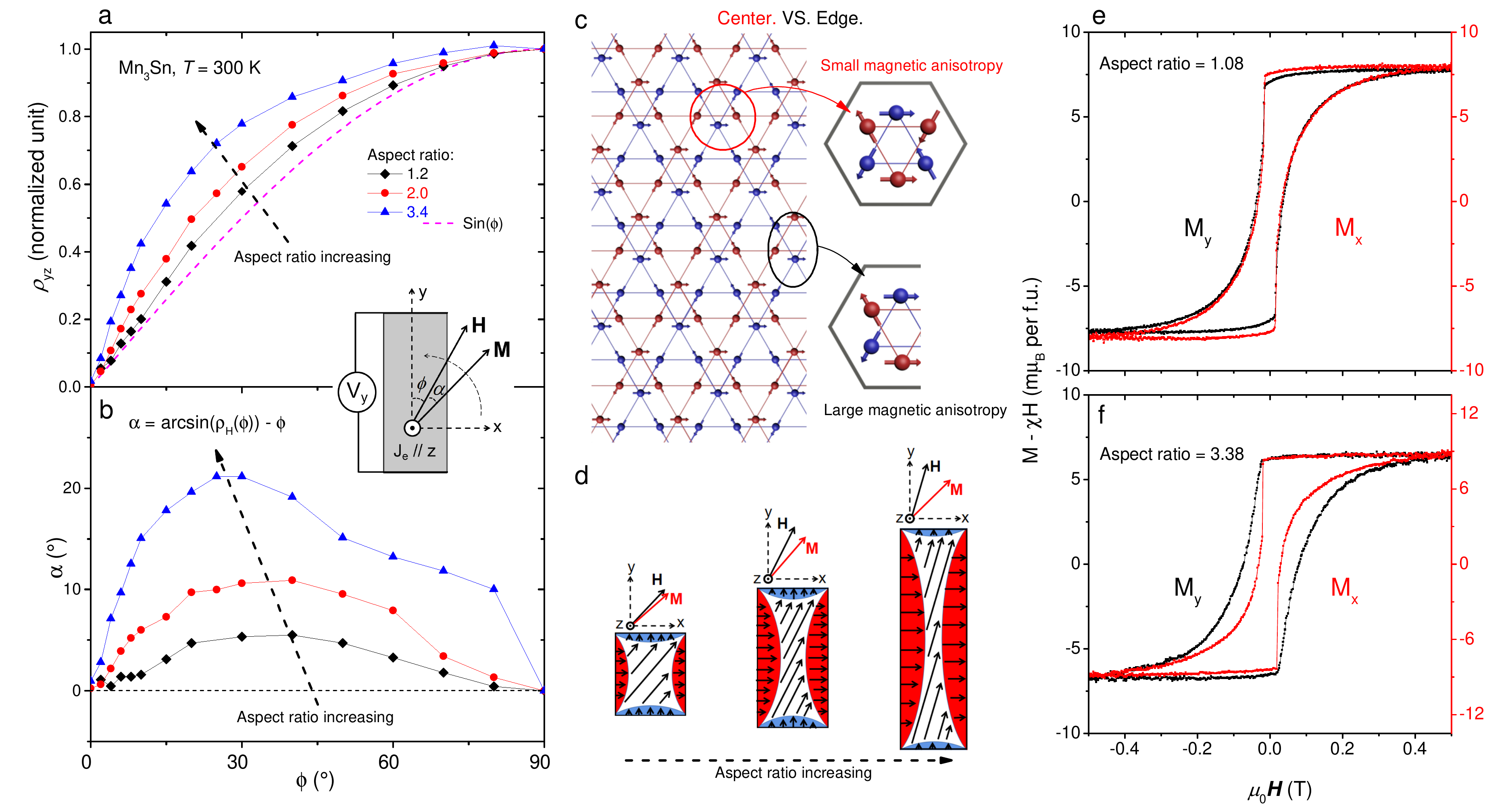}
  \caption{The dependence of the lag angle on aspect ratio: (a) variation of the amplitude of the Hall response\textcolor{red}{s} as a function of $\phi$, it is the angle between the magnetic field and the y-axis. Samples have different aspect ratios varying from 1.2 to 3.4. The experimental configuration and the normalization method are the same. (b) Angle dependence of the lag angle:  $\alpha = arcsin(\rho_{H}(\phi)) - \phi$. It steadily increases with the aspect ratio. (c) Sketch of the spin lattice at center and edge of xy-plane in Mn$_3$Sn crystal. The spin lattice at center of xy-plane exhibits a perfect octupole-like structure, which has high rotational symmetry and will present a small magnetic anisotropy. However, it is broken at edge of xy-plane, this will greatly damage the symmetry and generate a large uniaxial anisotropy. (d) Schematic diagram of the distribution of domains in samples with different aspect ratios. (e-f) Spontaneous magnetization in Mn$_3$Sn crystals with different aspect ratios \cite{Li2019}.}
  \label{fig: lag-angle}
\end{figure}

\quad Another insight is provided by the link between the aspect ratio of a sample and its lag angle ($\alpha$) \cite{ElBidweihy2017}, defined as:
\begin{equation}\label{lag angle}
\alpha = arcsin(\rho_{H}(\phi)) - \phi.
\end{equation}
Here $\phi$ is the angle between the magnetic field and the y-axis, $\rho_{H}(\phi)$ is the angular dependent Hall response. If the magnetization does not align along the orientation of the magnetic field, then there will be a finite angle between the total magnetization and the magnetic field. This lag angle is the source of the finite planar Hall effect. To verify this, we measured the angle dependence of the anomalous Hall effect in samples with different aspect ratios varying from 1.2 to 3.4, the results are shown in Fig.~\ref{fig: lag-angle}a. The angle-dependent Hall curves deviate from a sinusoidal behavior $sin(\phi)$. This deviation gradually increases with increasing aspect ratio. Given that the lag angle (See Fig.~\ref{fig: lag-angle}b) and the amplitude of PHE both increase with the enhancement of the aspect ratio, we conclude that the transverse magnetization, which emerges during domain reversal is intimately linked to the bulk-boundary dichotomy.  

\quad To explain this link, let us recall that spins in Mn$_3$Sn easily rotate in the xy-plane and the single-ion anisotropy \cite{Liu2017} is very small. The coupling between spins and lattice is so weak that in-plane rotation of magnetic field does not affect the amplitude of AHE~\cite{ Li2018}. The magnetic anisotropy energy, which is responsible to orient the magnetization along the easy axis, is remarkably small in Mn$_3$Sn. It has been quantified thanks to measurements of angle-dependent torque magnetometry~\cite{Duan2015}. At room temperature, the magnetic anisotropy energy in Mn$_3$Sn ($\approx 276 J/m^3$) is two orders of magnitude lower than  in Fe ($4.2 \times 10^4 J/m^3$) and three orders of magnitude lower than in Ni ($4.1 \times 10^5 J/m^3$)~\cite{Kittel}. The origin of this low magnetic anisotropy energy can be traced to the breathing Kagome spin texture. The octupole-like structure  has a high rotational symmetry and very weak ferromagnetism. The average magnetic moment in the ordered state is only 0.003 Bohr magneton per Mn atom. Now, at the edges of xy-plane, octupoles are broken and therefore the boundary induces a large uniaxial anisotropy as a result of the huge magnetic moment carried by an uncompensated Mn atom, which is few Bohr magnetons. Therefore, spins will resist rotation with the magnetic field at the boundary, much more than inside the bulk (See Fig.~\ref{fig: lag-angle}d). As a consequence, while at center of xy-plane, domains follow the orientation of the magnetic field, at the boundaries they remain pinned orthogonal to the edge. This provides a natural explanation for the observed dependence of the lag angle and the PHE on the aspect ratio. 

\quad This picture (See Fig.~\ref{fig: lag-angle}d) is also supported by an examination of the evolution of the magnetization hysteresis with aspect ratio. In a sample with $ar \approx 1$, the spontaneous magnetization is similar for two orthogonal directions (See Fig.~\ref{fig: lag-angle}e). In contrast, in a sample with a large $ar$, the jump in spontaneous magnetization is significantly sharper along the short side (See Fig.~\ref{fig: lag-angle}f). This is because the boundary orthogonal to the short side is dominant. Note that the magneto-crystalline anisotropy and the demagnetizing field (1 mT) ~\cite{Higo2018b} are too small in Mn$_3$Sn to provide a plausible explanation for such large lag angle and transverse magnetization.

\section{Discussion}
\quad These features have implications for optimizing the performance of our thermopile. Increasing the aspect ratio of the building bricks will enhance their transverse magnetization and provide higher stability. The advantage of the present design is its simple structure, which makes its assembly possible by an one-step sputtering method. More importantly, the leg contacts of the traditional thermopile  is avoided.

\quad However, the Nernst thermopower of Mn$_3$Sn used is low ($\sim$ 0.5 $\mu$V/K \cite{Li2017}). This is two orders of magnitude lower than the Seebeck thermopower of Bi$_2$Te$_3$ (or  Bi$_2$Se$_3$)  of a Seebeck thermopile. This is a setback. Nevertheless, the design is promissing for a number of reasons.

\quad First of all, our design allows a larger density of bricks. For a heat flux sensor\cite{Zhou2020b}, what matters is the overall sensitivity of the array and not just the magnitude of thermopower.

\quad Moreover, the idea behind this work can be used  to make thermopiles  with other magnetic materials with a larger ANE, known \cite{Sakai2020, Sakuraba2016} or yet to be discovered \cite{Xu2020b}. Ferromagnets with strong magneto-crystalline anisotropy or shape anisotropy could be candidates. In this context, let us notice that Zhou \textit{et al.} \cite{Zhou2021} have recently reported  on a Seebeck-driven transverse thermoelectric generation  leading to a signal as large as 82.3 (-41.0) $\mu$V/K  in a hybrid material Co$_2$MnGa/n(p)-type-Si. This is one order of magnitude larger than the record-high S$_{ANE}$.

\section{Conclusion}
\quad In summary, we proposed and realized a new mono-material Nernst thermopile with hermaphroditic legs. The design has a number of obvious advantages compared to previous proposals for thermopiles. While the  magnitude of the Nernst thermopower in Mn$_3$Sn is too low, the idea can be used with other magnetic materials with a larger Nernst thermopower. 

\section{Experimental Section}
\threesubsection{Single crystal growth}\\ \quad Large Mn$_3$Sn single crystals were grown by a two-step method \cite{Li2017, Li2018}. Firstly, the raw materials (99.999\% Mn, 99.999\% Sn) with the molar ratio of 3.3 : 1 were weighted and mixed for the precursor crystal growth. The raw materials were heated up to 1100 $^{\circ}$C, remained there for 2 hour to ensure homogeneity of the melting mixture, and were cooled down slowly to 900$^{\circ}$C for the precursor crystal growth. Then, the precursor crystal ingots were ground and put in an alumina crucible, which was sealed in a quartz tube and hung in a vertical Bridgman furnace. The crucible was slowly cooled below the highest temperature point (1050$^{\circ}$C) for the single crystal growth. This was repeated three times with different rates, 2 mm/h, 2 mm/h and 1 mm/h, in order to purify the single crystal. Finally, the obtained single crystal was annealed at 850$^{\circ}$C for 20 hours and then quenched to room temperature. Using energy dispersive X-ray spectroscopy (EDX), the stoichiometry of Mn$_3$Sn was found to be Mn$_{3.22}$Sn, this is close to but slightly below the ratio of the raw materials.

\threesubsection{Sample preparation}\\ \quad Large Mn$_3$Sn single crystals were cut to desired dimensions by a wire saw. Six crystals with a similar size of $\simeq 0.88 \times 0.26 \times 0.2 mm^3$  were cut from the same mother crystal. They were fixed on a Al$_3$O$_2$ heat sink using GE Vanish. Pairs of crystals forming a bicrystal were connected together by the sliver paste. A 1000 $\Omega$  resistor chip on a copper flake was used as a heater. 

\threesubsection{Measurement}\\ \quad All transport experiments were performed in a commercial Quantum Design PPMS, with the Horizontal Rotator Option and the DC resistivity puck. Hall resistivity was measured by a standard four-probe method using a current source (Keithley6221) and a DC-nanovoltmeter (Keithley2182A). For  Nernst measurements, two Chromel- Constantan (type E) thermocouples were employed to measure the temperature difference inside a high-vacuum environment.

\medskip

\medskip
\textbf{Acknowledgements} \par 
\quad X. L. acknowledges Dongwang Yang for stimulating discussions. This work was supported by the National Science Foundation of China (Grant No. 11574097 and No. 51861135104) and The National Key Research and Development Program of China (Grant No.2016YFA0401704). This work was supported in France by the Agence Nationale de la Recherche (ANR-18-CE92-0020-01; ANR-19-CE30-0014-04). X. L. acknowledges the China National Postdoctoral Program for Innovative Talents (Grant No.BX20200143) and the China Postdoctoral Science Foundation (Grant No.2020M682386).

\medskip

%
\bibliographystyle{MSP}
\bibliography{MSP-template}

\begin{thebibliography}{99}

\bibitem{Liu2015} W. Liu, Q. Jie, H. S. Kim, Z. Ren, \textit{Acta Materialia} \textbf{2015}, \textit{87}, 357.

\bibitem{He2017} J. He, T. M. Tritt \textit{Science} \textbf{2017}, \textit{357}, 6358.

\bibitem{He2018} R. He, G. Schierning, K. Nielsch, \textit{Adv. Mater. Technol.} \textbf{2018}, \textit{3}, 1700256.

\bibitem{Niemann2016} A. C. Niemann, T. B\"ohnert, A. K. Michel, S. B\"a$\beta$ler, B. Cotsmann, K. Neur\'ohr, B. T\'oth, L. P\'eter, I. Bakonyi, V. Vega, V. M. Prida, J. Gooth, K. Nielsch, \textit{Adv. Electron. Mater.} \textbf{2016}, \textit{2}, 1600058.

\bibitem{Behnia2016} K. Behnia, H. Aubin, \textit{Rep. Prog. Phys.} \textbf{2016}, \textit{79}, 046502.

\bibitem{Harman1964} T. C. Harman, J. M. Honig, S. Fischler, A. E. Paladino, M. J. Button, \textit{Appl. Phys. Lett.} \textbf{1964}, \textit{4}, 77.

\bibitem{Zhou2021} W. Zhou, K. Yamamoto, A. Miura, R. Lguchi, Y. Miura, K. I. Uchida, Y. Sakuraba, \textit{Nat. Mater.} \textbf{2021}, doi:10.1038/s41563-020-00884-2.

\bibitem{Sakuraba2013} Y. Sakuraba, K. Hasegawa, M. Mizuguchi, T. Kubota, S. Mizukami, T. Miyazaki, K. Takanashi, \textit{Appl. Phys. Express} \textbf{2013}, \textit{6}, 033003.

\bibitem{Sakai2020} A. Sakai, S. Minami, T. Koretsune, T. Chen, T. Higo, Y. Wang. T. Nomoto, M. Hirayama, S. Miwa, D. Nishio-Hamane, F. Ishii, R. Arita, S. Nakatsuji, \textit{Nature} \textbf{2020}, \textit{581}, 53.


\bibitem{Sakuraba2016} Y. Sakuraba, \textit{Scri. Mater.} \textbf{2016}, \textit{111}, 29.

\bibitem{Yang2017a} Z. Yang, E. A.  Codecido, J. Marquez, Y. Zheng, J. P. Heremans, R. C.  Myers, \textit{AIP Adv.} \textbf{2017}, \textit{7}, 095017.

\bibitem{Mizuguchi2019} M. Mizuguchi, S. Nakatsuji, \textit{Sci. Technol. Adv. Mater.} \textbf{2019}, \textit{20}, 262.

\bibitem{Ikhlas2017} M. Ikhlas, T. Tomita, T. Koretsune, M. T. Suzuki, D. Nishio-Hamane, R. Arita. Y. Otani, S. Nakatsuji, \textit{Nat. Phys.} \textbf{2017}, \textit{13}, 1085.


\bibitem{Zhou2020a} X. Zhou, J. P. Hanke, W. Feng, S. Bl\"ugel, Y, Mokrousov, Y. Yao, \textit{Phy. Rev. Mater.} \textbf{2020}, \textit{4}, 024408.

\bibitem{Fu2020} C. Fu, Y. Sun, C. Felser, \textit{APL Mater.} \textbf{2020}, \textit{8}, 040913.

\bibitem{Xiao2006} D. Xiao, Y. Yao, Z. Fang, Q. Niu, \textit{Phys. Rev. Lett.} \textbf{2006}, \textit{97}, 026603.

\bibitem{Li2017} X. Li, L. Xu, L. Ding, J. Wang, M. Shen, X. Lu, Z. Zhu, K. Behnia, \textit{Phys. Rev. Lett.} \textbf{2017}, \textit{119}, 056601.

\bibitem{Akito2018} A. Sakai, Y. P. Mizuta, A. A. Nugroho, R. Sihombing, T. Koretsune, M. Suzuki, N. Takemori, R. Ishii, D. Nishio-Hamane, R. Arita, P. Goswami, S. Nakatsuji, \textit{Nat. Phys.} \textbf{2018}, \textit{14}, 1119.

\bibitem{Guin2019} S. N. Guin, P. Vir, Y. Zhang, N. Kumar, S. J. Watzman, C. Fu, E. Liu, K. Manna, W. Schnelle, J. Gooth, C. Shekhar, Y. Sun, C. Felser, \textit{Adv. Mater.} \textbf{2019}, \textit{31}, 1806622.

\bibitem{Ding2019} L. Ding, J. Koo, L. Xu, X. Li, X. Lu, L. Zhao, Q. Wang, Q. Yin, H. Lei, B. Yan, Z. Zhu, K. Behnia, \textit{Phys. Rev. X.} \textbf{2019}, \textit{9}, 041061.

\bibitem{Xu2019} J. Xu, W. A. Phelan, C. L. Chien, \textit{Nano. Lett.} \textbf{2019}, \textit{19}, 8250.

\bibitem{Xu2020} L. Xu, X. Li, X. Lu, C. Collignon, H. Fu, J. Koo, B. Fauqu\'e, B. Yan, Z. Zhu, K. Behnia, \textit{Sci. Adv.} \textbf{2020}, \textit{6}, eaaz3522.

\bibitem{Narita2017} H. Narita, M. Ikhlas, M. Kimata, A. A. Nugroho, S. Nakatsuji, Y. Otani, \textit{Appl. Phys. Lett.} \textbf{2017}, \textit{111}, 202404.

\bibitem{Zimmer1972} G. J. Zimmer, E. Kr\'{e}n, \textit{AIP Conf. Proceed.} \textbf{1972}, \textit{5}, 513.

\bibitem{Tomiyoshi1982} S. Tomiyoshi, \textit{J. Phys. Soc. Jpn.} \textbf{1982}, \textit{51}, 803.

\bibitem{Tomiyoshi1982b} S. Tomiyoshi, Y. Yamaguchi, \textit{J. Phys. Soc. Jpn.} \textbf{1982}, \textit{51}, 2478.

\bibitem{Yang2017b} H. Yang, Y. Sun, Y. Zhang, W. J. Shi, S. S. P. Parkin, B. Yan, \textit{New J. Phys.} \textbf{2017}, \textit{19}, 015008.

\bibitem{Nakatsuji2015}  S. Nakatsuji, N. Kiyohara, T. Higo, \textit{Nature} \textbf{2015}, \textit{527}, 212.

\bibitem{Higo2018a} T. Higo, D. Qu, Y. Li, C. L. Chien, Y. Otani, S. Nakatsuji, \textit{Appl. Phys. Lett.} \textbf{2018}, \textit{113}, 202402.

\bibitem{Higo2018b} T. Higo, H. Man, D. B. Gopman, L. Wu, T. Koretsune, O. M. J. Van't Erve, Y. P. Kabanov, D. Rees, Y. Li, M. T. Suzuki, S. Patankar, M. Ikhlas, C. L. Chien, R. Arita, R. D. Shull, J. Orenstein, S. Nakatsuji, \textit{Nat. Photonics.} \textbf{2018}, \textit{12}, 73.

\bibitem{Li2019} X. Li, C. Collignon, L. Xu, H. Zuo, A. Cavanna, U. Gennser, D. Mailly, B. Fauqu\'e, L. Balents, Z. Zhu, K. Behnia, \textit{Nat. Commun.} \textbf{2019}, \textit{10}, 3021.

\bibitem{Suzuki2017} M. T. Suzuki, T. Koretsune, M. Ochi, R. Arita, \textit{Phys. Rev. B} \textbf{2017}, \textit{95}, 094406.

\bibitem{ElBidweihy2017} H. ElBidweihy, \textit{IEEE Trans. Magn.} \textbf{2017}, \textit{53}, 11.

\bibitem{Liu2017} J. Liu, L. Balents, \textit{Phys. Rev. Lett.} \textbf{2017}, \textit{119}, 087202.

\bibitem{Li2018} X. Li, L. Xu, H. Zuo, A. Subedi, Z. Zhu, K. Behnia, \textit{SciPost Phys.} \textbf{2018}, \textit{5}, 063.

\bibitem{Duan2015} T. F. Duan, W. J. Ren, W. L. Liu, S. J. Li, W. Liu, Z. D. Zhang, \textit{Appl. Phys. Lett.} \textbf{2015}, \textit{107}, 082403.

\bibitem{Kittel} C. Kittel, \textit{Introduction to the Solid state physics.} John Wiley \& Sons, New York \textbf{1966}.

\bibitem{Zhou2020b} W. Zhou, Y. Sakuraba, \textit{Appl. Phys. Express.} \textbf{2020}, \textit{13}, 043001.


\bibitem{Xu2020b} L. Xu, X. Li, L. Ding, T. Chen, A. Sakai, B. Fauque, S. Nakatsuji, Z. Zhu, K. Behnia, \textit{Phys. Rev. B} \textbf{2020}, \textit{101}, 180404(R).




\end{thebibliography}



\end{document}